\documentclass[12pt]{article}
\usepackage{amsfonts,amsmath}
\usepackage{amssymb}

\usepackage{graphicx}

\makeatletter \@addtoreset{equation}{section} \makeatother

{\vspace{3mm} }

\def\*{\star}

\def\E2{\mathbf{E}}

\newcommand{\be}{\begin{equation}}
\newcommand{\ee}{\end{equation}}
\newcommand{\bee}{\begin{eqnarray}}
\newcommand{\beee}{\begin{array}}
\newcommand{\eee}{\end{eqnarray}}
\newcommand{\eeee}{\end{array}}
\newcommand{\nuk}{\nu\mathcal{K}}

\newcommand{\Aq}{\mathsf{Aq}}

\begin{document}

\begin{flushright}
FIAN/TD/15-2020\\
\end{flushright}

\vspace{0.5cm}
\begin{center}
{\large\bf Star product for deformed oscillator algebra $\Aq(2,\nu)$}

\vspace{1 cm}

\textbf{A.V.~Korybut}\\

\vspace{1 cm}

\textbf{}\textbf{}\\
 \vspace{0.5cm}
 \textit{ I.E. Tamm Department of Theoretical Physics,
Lebedev Physical Institute,}\\
 \textit{ Leninsky prospect 53, 119991, Moscow, Russia }\\

\par\end{center}

\begin{center}
\vspace{0.6cm}
% didenko@lpi.ru, vasiliev@lpi.ru \\

\par\end{center}

\vspace{0.4cm}

\begin{abstract}
\noindent An analogue of the Moyal star product is presented for the  deformed oscillator algebra. It contains several homotopy-like additional integration parameters in the
 multiplication kernel generalizing the differential Moyal star-product formula
 $\exp[i\epsilon_{\alpha\beta}\partial^\alpha \partial^\beta]$. Using Pochhammer formula
 \cite{Pochhammer},
 integration over these parameters is carried over a Riemann surface associated with the expression of the type $z^x (1-z)^y$ where $x$ and $y$ are arbitrary real numbers.
\end{abstract}

\section{Introduction}

Possible deformations of standard oscillator commutation relations
\begin{equation}
[y_\alpha,y_\beta]=2i\epsilon_{\alpha \beta},\;\; \alpha,\beta=\overline{1,2}
\end{equation}
that nonetheless lead to equally spaced energy spectrum of deformed oscillator was studied by Wigner \cite{Wigner}. He found that there should be one-parameter family of deformed commutation relations. With the help of additional anticommuting operator Wigner's deformed oscillator algebra can be represented  in the form \cite{Vasiliev1}
\begin{equation}\label{commut}
[y_\alpha,y_\beta]=2i \epsilon_{\alpha \beta}(1 +\nu \mathcal{K}),\;\; \lbrace y_\alpha, \mathcal{K} \rbrace =0,\;\; \mathcal{K}^2=1.
\end{equation}
Here $\nu \in \mathbb{C}$ is an arbitrary parameter and $\mathcal{K}$ is the so called Klein operator. $\Aq(2,\nu)$ is the associative algebra generated as universal enveloping algebra of
these (anti)commutaton relations. Generic element of $\Aq(2,\nu)$ can be written in the form of formal power series as
\begin{equation}
f(y,\mathcal{K})=\sum_{n=0}^{\infty}\sum_{A=0}^{1} f^{\alpha_1\dots \alpha_n}_A y_{\alpha_1}\dots y_{\alpha_n} \mathcal{K}^A,
\end{equation}
where tensors $f^{\alpha_1\dots \alpha_n}_A$ are totally symmetric in upper indices. From now on Weyl ordering of oscillators is assumed. Product of two generic elements should be again written as formal power series of oscillators contracted with totally symmetric in upper indices coefficients, i.e.
\begin{equation}\label{genProd}
f(y,\mathcal{K})\ast g(y,\mathcal{K})=h(y,\mathcal{K})=\sum_{n=0}^{\infty}\sum_{A=0}^{1} h^{\alpha_1\dots \alpha_n}_A y_{\alpha_1}\dots y_{\alpha_n} \mathcal{K}^A.
\end{equation}
Here symmetrization is supposed to be performed using (anti)commutation relations \eqref{commut}. To compute r.h.s. of \eqref{genProd} one should use structure constants $\mathsf{H}(m,n,p,\nu \mathcal{K})$ for monomials found in \cite{Korybut}
\begin{multline}\label{prod1}
f^{\alpha_1 \dots \alpha_m}_A y_{\alpha_1}\dots y_{\alpha_m} \mathcal{K}^A \ast g^{\beta_1\dots \beta_n}_B y_{\beta_1}\dots y_{\beta_n} \mathcal{K}^B=\\
=f^{\alpha_1 \dots \alpha_m}_A y_{\alpha_1}\dots y_{\alpha_m}  \ast g^{\beta_1\dots \beta_n}_B y_{\beta_1}\dots y_{\beta_n} (-1)^{n A} \mathcal{K}^A \mathcal{K}^B=\\
=f^{\alpha_1 \dots \alpha_m}_A g^{\beta_1\dots \beta_n}_B \sum_{p=0}^{\min(m,n)}  i^p \epsilon_{\alpha_1 \beta_1}\dots \epsilon_{\alpha_p \beta_p} y_{(\alpha_1}\dots y_{\alpha_{m-p}}y_{\beta_1}\dots y_{\beta_{n-p})} \mathsf{H}(m,n,p,\nu \mathcal K) (-1)^{nA}\mathcal{K}^{A+B},
\end{multline}
where oscillators on the r.h.s. are totally symmetrized, i.e.
\begin{equation}
y_{(\alpha_1}\dots y_{\alpha_{m-p}}y_{\beta_1}\dots y_{\beta_{n-p})}=\frac{1}{(m+n-2p)!}(y_{\alpha_1}\dots y_{\alpha_{m-p}}y_{\beta_1}\dots y_{\beta_{n-p}}+\text{all permutations}).
\end{equation}
Structure constants depend on parities of monomials in the product. Explicit formulas for each of parities are given in the next section.

(Anti)commutation relations \eqref{commut} play the important role in HS theory since they determine the form of the full nonlinear system of equations that acquires HS symmetry as a gauge symmetry of the theory \cite{Vasiliev3_6},\cite{Vasiliev4}. Moreover in case of 2+1 dimensions nonlinear system can be naturally reformulated in terms of deformed oscillators \cite{Prokushkin}.

In purely two dimensional conformal theory context the algebra was studied in \cite{Pope},\cite{Berg2} and \cite{Berg3}. Development of $AdS_3/CFT_2$ correspondence showed importance of deformed oscillator algebra from another perspective \cite{Gaberdiel},\cite{KoreanGuy}. In $AdS_3/CFT_2$ correspondence algebras $\mathfrak{hs}[\lambda]$ and $\mathfrak{shs}[\lambda]$ turn out to be important. They are the (super)Lie algebras constructed by taking the quotient of universal enveloping algebras $\mathcal{U}(\mathfrak{sp}(2))$ and $\mathcal{U}(\mathfrak{osp}(2\vert 1))$ over the ideal generated by the quadratic Casimir of $\mathfrak{sp}(2)$ \cite{Feigin} and $\mathfrak{osp}(2\vert 1)$ \cite{Berg3}, respectively.

For the first time deformed oscillator algebra was argued to be interpreted as a higher spin algebra in \cite{Vasiliev1},\cite{Vasiliev2}. In a slightly different realization higher spin algebra was studied in \cite{Berg1}. The associative algebra underlying  $\mathfrak{hs}[\lambda]$ and $\mathfrak{shs}[\lambda]$ is $\Aq(2,\nu)$ restricted to the case of even powers in $y$ or unrestricted, respectively. The associative product underlying $\mathfrak{hs}[\lambda]$ was for the first time introduced in \cite{Pope}. However associativity of this product (Lone-Star product as it was called by the authors) was only conjectured, the proof was given in \cite{Korybut}.  In \cite{Joung} product for two even in $y$ functions of very specific type was given. $\mathfrak{shs}[\lambda]$ from deformed oscillators point of view was considered in \cite{Boulanger}. Algebra $\Aq(2,\nu)$ in a different from the deformed oscillators point of view was studied in \cite{Fradkin}.

There is also another approach to higher spin (super)algebras called factorization by projector \cite{HSsuperalgebra}, \cite{HSsuperalgebra1}, \cite{HSsuperalgebra3}. The main idea is to find projector $\Delta$\footnote{Technically $\Delta$ found in \cite{HSsuperalgebra} is not the projector since $\Delta^2$ diverges} that acts trivially on the elements from the ideal
\begin{equation}
\Delta a=a\Delta=0,\;\;\; a\in \mathcal{I}.
\end{equation}
Then product in the quotient algebra can be written in the form
\begin{equation}
g_1 \circ g_2=f_1 f_2 \Delta,\;\; g_i=f_i \Delta.
\end{equation}
Equation for the projector for $\mathfrak{hs}[\lambda]$ case was found in \cite{Alkalaev}.

In the absence of deformation ($\nu=0$), the product of two elements can
 be written in the well-known Moyal form
\begin{equation}\label{Moyal}
f(y)\ast g(y)=f(y)\sum_{p=0}^\infty \frac{i^p}{p!} \left(\frac{\overleftarrow{\partial}}{\partial y_\alpha}\epsilon_{\alpha \beta} \frac{\overrightarrow{\partial}}{\partial y_\beta}\right)^p g(y).
\end{equation}
Main result of this paper is the analogue of the Moyal product for the generating commutation relations  \eqref{commut}.

The paper is organized as follows: in section 2  structure constants for the associative product in $\Aq(2,\nu)$ are presented, in the section 3  difficulties of integral representation are discussed and possible way to overcome them with the help of Pochhammer representation for Euler beta-function is presented, in section 4 star product for generic formal power series is given. Conclusion contains discussion of the obtained result and future directions.

\section{Structure constants for $\Aq(2,\nu)$}\label{Constants}
Structure constants $\mathsf{H}(m,n,p,\nu \mathcal{K})$ in \eqref{prod1} obtained in \cite{Korybut} depend on the parity of product factors, i.e.
\begin{equation}\label{constants}
\mathsf{H}(m,n,p,\nu \mathcal{K})=\begin{cases}
A(m,n,p,\nu\mathcal{K})\, ,\,\, m\,\,\, \text{is even}\, , n\, \,\, \text{is even}\, ,\\
B(m,n,p,\nu\mathcal{K})\, ,\,\, m\,\,\, \text{is odd}\, ,\; n\, \,\, \text{is odd}\, ,\\
C(m,n,p,\nu\mathcal{K})\, ,\,\, m\,\,\, \text{is even}\, , n\, \,\, \text{is odd}\, ,\\
D(m,n,p,\nu\mathcal{K})\, ,\,\, m\,\,\, \text{is odd}\, ,\; n\, \,\, \text{is even}\, .
\end{cases}
\end{equation}
Since all structure constants in \eqref{constants} are expressible in terms of $A(m,n,p,\nu\mathcal{K})$  in the sequel of this section we assume that $m$ and $n$ are even
\begin{equation}\label{A}
A(m,n,p,\nu\mathcal{K})=\frac{i^p\, m!n!}{(m-p)!(n-p)!p!} {}_4 F_3\left[\begin{matrix} 1-\frac{\nuk}{2}& \frac{\nuk}{2} &\frac{-p}{2}& \frac{1-p}{2}\\
\frac{1-m}{2}& \frac{1-n}{2} &\frac{m+n-2p+3}{2}\end{matrix}\Bigg|1\right].
\end{equation}
Here ${}_4 F_3[\dots]$ is the generalized hypergeometric function
\begin{equation}\label{hyper}
{}_4 F_3 \left[\begin{matrix}
a & b & c & d\\
e & f & g
\end{matrix}\Bigg| z\right]=\sum_{q=0}^\infty\frac{(a)_q (b)_q (c)_q (d)_q}{(e)_q (f)_q (g)_q}\frac{z^q}{q!}\, ,
\end{equation}
where $(a)_q$ is the so called descending Pochhammer symbol defined as
\begin{equation}\label{symb}
(a)_q\equiv \frac{\Gamma(a+q)}{\Gamma(a)}.
\end{equation}
\begin{multline}\label{B}
B(m+1,n+1,p,\nuk)=A(m,n,p,-\nuk)+i(m+n-2p+3+\nuk)A(m,n,p-1,-\nuk)+\\
+i^2 (m+2-p)(n+2-p)\frac{m+n-2p+3\nuk}{m+n-2p+3}\frac{m+n-2p+5\nuk}{m+n-2p+3}A(m,n,p-2,-\nuk),
\end{multline}
\begin{multline}\label{C}
C(m,n+1,p,\nuk)=A(m,n,p,-\nuk)+\\
+i(m+1-p)\frac{m+n-2p+3+\nuk}{m+2-2p+3}A(m,n,p-1,-\nuk),
\end{multline}
\begin{equation}\label{D}
D(m+1,n,p,\nuk)=A(m,n,p,\nuk)+i(n+1-p)\frac{m+n-2p+3-\nuk}{m+n-2p+3}A(m,n,p-1,\nuk).
\end{equation}
Let us stress that structure constants \eqref{A},\eqref{B},\eqref{C},\eqref{D} are unambiguously determined by (anti)commutation relations \eqref{commut} and associativity of algebra $\Aq(2,\nu)$.

Using structure constants one can multiply any formal power series. But suppose one has to solve an equation of the form
\begin{equation}\label{EQ}
f(y)\ast X(y)=g(y)
\end{equation}
where function $f$ and $g$ are known and $X$ is to be defined. If there is an integral representation of the product as
\begin{equation}
f(y)\ast X(y)= \int d\tau f(\tau y)\left[\sum_{p=0}^\infty\frac{i^p}{p!}\left(\frac{\overleftarrow{\partial}}{\partial y_\alpha}\epsilon_{\alpha \beta} \frac{\overrightarrow{\partial}}{\partial y_\beta}\right)^p K(\tau,p)\right]X(\tau y),
\end{equation}
then equation \eqref{EQ} can be turned into integro-differential equation which might be easier to solve. With the help of Pochhammer representation for beta-function such representation for the product is constructed.

\section{Pochhammer representation for structure constants}
\subsection{Even $\times$ even case}
For positive values of $a,b,c,d,e,f,g$ and when upper arguments are bounded by lower ones hypergeometric function \eqref{hyper} admits integral representation due to integral representation for Euler beta-function, i.e.
\begin{eqnarray}\label{Euler}
B(x,y)=\int_0^1 dt\, t^{x-1} (1-t)^{y-1}\; ,\; x>0\, ,\, y>0\,.
\end{eqnarray}
For example if $d>g>0$ and $z\in \mathbb{R}$ one can express ${}_4F_3$ as an integral of ${}_3F_2$ 
\begin{equation}\label{EulerHyper}
{}_4 F_3\left[\left. \begin{matrix} a & b & c & d \\
e & f & g \end{matrix}\right|  z \right]=\frac{\Gamma(g)}{\Gamma(d)\Gamma(g-d)}\int_0^1 dt\, t^{d-1}(1-t)^{g-d-1} {}_3F_2\left[\left. \begin{matrix}
a & b & c\\
e & f \end{matrix} \right| tz \right]
\end{equation}
One can proceed further and express ${}_3F_2$ as an integral of ${}_2F_1$ if there is another positive pair of lower and upper arguments where upper one is bounded by the lower one.
%If $b_k > a_k> 0$ and $z\in \mathbb{R}$ the following formula is valid
%\begin{multline}
%{}_4 F_3\left[\left. \begin{matrix} a_1 & a_2 & a_3& a_4 \\
%b_1& b_2 &b_3 \end{matrix}\right|  z \right]=\\
%=\left(\prod_{k=1}^{3}\frac{\Gamma(b_k)}{\Gamma(a_k)\Gamma(b_k-a_k)}\right)\int_0^1 dt_1\int_0^1 dt_2 \int_0^1 dt_3\, \left(\prod_{j=1}^3 t_j^{a_j-1}(1-t_j)^{b_j-a_j-1}\right)\left(1-zt_1 t_2 t_3\right)^{-a_4}
%\end{multline}

However, since some arguments of hypergeometric function \eqref{A} that depend on powers of the product factor monomials are negative,  standard formula like \eqref{EulerHyper} is not applicable
\begin{equation}
{}_4 F_3\left[\left. \begin{matrix} 1-\frac{\nuk}{2}& \frac{\nuk}{2} &\frac{-p}{2}& \frac{1-p}{2}\\
\underbrace{{\frac{1-m}{2}}}_{<0}& \underbrace{\frac{1-n}{2}}_{<0} &\frac{m+n-2p+3}{2}\end{matrix}\right|  1\right].
\end{equation}

\begin{figure}[h]\label{Contour}
\begin{center}
\includegraphics[width=10cm]{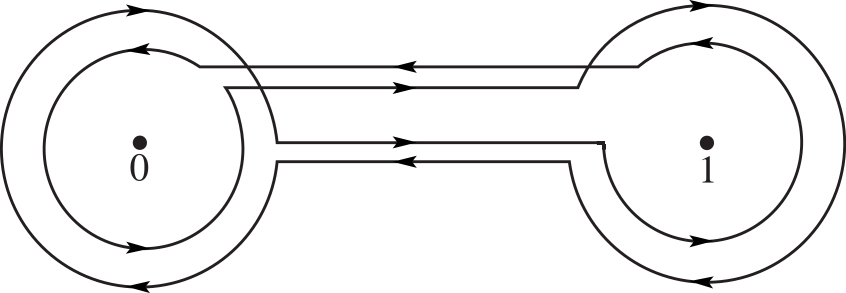}.
\end{center}
\caption{Pochhammer contour}
\end{figure}

However there is a remarkable alternative representation for Euler beta-function, the so called Pochhammer representation \cite{Pochhammer}
\begin{equation}\label{Poch}
\int_C dz \, z^{x-1}\left(1-z\right)^{y-1}=\left(1-e^{2\pi i x}\right)\left(1-e^{2\pi i y}\right)B\left(x,y\right),
\end{equation}
where integration is carried on the Riemann surface defined by the integrand along the contour on Fig.~\ref{Contour}.
%\begin{center}
%\includegraphics[width=10cm]{Pochhammer1.png}.
%\end{center}

\noindent Moreover formula \eqref{Poch} gives analytic continuation of Euler beta-function to the complex plane ($x,y \in \mathbb{C}$). Even though the last lower argument of hypergeometric function, namely ($\frac{m+n-2p+3}{2}$), is positive there is no upper positive argument that fulfills requirements for using Euler representation \eqref{Euler}. I.e. real  part of upper argument should be bounded by $\frac{m+n-2p+3}{2}$ which is not the case for generic values of $\nu$.

Because of the phase factors on the r.h.s. of \eqref{Poch} one should use Pochhammer representation twice. Indeed,
suppose we want to obtain $\left[\left(\frac{1-m}{2}\right)_q\right]^{-1}$ which is by definition
\begin{equation}\label{sample}
\left[\left(\frac{1-m}{2}\right)_q\right]^{-1}=\frac{\Gamma\left(\frac{1-m}{2}\right)}{\Gamma\left(\frac{1-m}{2}+q\right)}.
\end{equation}
Consider an integral
\begin{equation}\label{int1}
I_1=\int_C dz\, z^{\frac{1-m}{2}-\xi-1}(1-z)^{q+\xi-1}=-2i\sin(2\pi \xi)\frac{\Gamma\left(\frac{1-m}{2}-\xi\right)\Gamma\left(q+\xi\right)}{\Gamma\left(\frac{1-m}{2}+q\right)}.
\end{equation}
Here $\xi$ is some non-integer number and the phase factor was simplified because $m$ is even and $q$ is integer. Note that r.h.s. of \eqref{int1} contains the same Gamma-function in denominator as r.h.s. of \eqref{sample}. To obtain proper Gamma-function in enumerator consider the following integral
\begin{equation}\label{int2}
I_2=\int_C dz\, z^{\frac{1-m}{2}-1}(1-z)^{-\xi-1}=2\left(1-e^{-2\pi i \xi}\right)\frac{\Gamma\left(\frac{1-m}{2}\right)\Gamma\left(-\xi\right)}{\Gamma\left(\frac{1-m}{2}-\xi\right)}
\end{equation}
Product of \eqref{int1} and \eqref{int2} gives
\begin{equation}\label{pre1}
I_1I_2=-4i\left(1-e^{-2\pi i \xi}\right)\sin\left(2\pi \xi\right)\Gamma\left(\xi\right)\Gamma\left(-\xi\right)\frac{\left(\xi\right)_q}{\left(\frac{1-m}{2}\right)_q }.
\end{equation}
Using the following Gamma-function identities 
\begin{equation}\label{GammaIdentities}
\Gamma\left(1-\xi\right)=(-\xi)\Gamma(-\xi)\; ,\;\; \Gamma\left(\xi\right)\Gamma(1-\xi)=\frac{\pi}{\sin\left(\pi \xi\right)}
\end{equation}
the prefactor can be simplified and expression \eqref{pre1} turns into
\begin{equation}\label{pre2}
I_1I_2=-\frac{8\pi}{\xi} \sin\left(2\pi\xi\right)e^{-i\pi \xi}\frac{\left(\xi\right)_q}{\left(\frac{1-m}{2}\right)_q }.
\end{equation}

Analogously one can represent $\left[\left(\frac{1-m}{2}\right)_q\right]^{-1}$ and $\left[\left(\frac{m+n-2p+3}{2}\right)_q\right]^{-1}$ introducing new non-integer variables $\eta$ and $\zeta$ respectively.
To reproduce structure constants \eqref{A} we define following functions
\begin{equation}\label{calF}
\mathcal{F}\left(p,\nuk,s_1,t_1,u_1\right)={}_4F_3\left[\left.\begin{matrix}
1-\frac{\nuk}{2} & \frac{\nuk}{2} & -\frac{p}{2} & \frac{1-p}{2}\\
\xi & \eta &\zeta
\end{matrix} \right|(1-s_1)(1-t_1)(1-u_1)\right],
\end{equation}
\begin{multline}
\mathcal{R}(s_1,s_2,t_1,t_2,u_1,u_2)=s_1^{-\frac{1}{2}-\xi}s_2^{-\frac{1}{2}}\frac{(1-s_1)^{\xi-1}}{(1-s_2)^{\xi+1}} t_1^{-\frac{1}{2}-\eta}t_2^{-\frac{1}{2}}\frac{(1-t_1)^{\eta-1}}{(1-t_2)^{\eta+1}}u_1^{\frac{1}{2}-\zeta}u_2^{\frac{1}{2}}\frac{(1-u_1)^{\zeta-1}}{(1-u_2)^{\zeta+1}}.
\end{multline}
Also we introduce shorthand notation for the integrals
\begin{equation}
\int d\Gamma \equiv \int_{C_{s_1}} ds_1\int_{C_{s_2}} ds_2\int_{C_{t_1}} dt_1\int_{C_{t_2}} dt_2\int_{C_{u_1}} du_1\int_{C_{u_2}} du_2,
\end{equation}
where integration contours are Pochhammer contours (Fig~\ref{Contour}). To see how structure constants are reproduced consider the following expression
\begin{multline}\label{master}
\int d\Gamma \left(\sqrt{\frac{u_1 u_2}{s_1 s_2}}\right)^m \mathcal{R}(s_1,s_2,t_1,t_2,u_1,u_2)\mathcal{F}(p,\nuk,s_1,t_1,u_1)\frac{1}{(u_1 u_2)^p}\left(\sqrt{\frac{u_1u_2}{t_1 t_2}}\right)^n=\\
=\sum_{q=0}^\infty\frac{\left(1-\frac{\nuk}{2}\right)_q\left(\frac{\nuk}{2}\right)_q\left(-\frac{p}{2}\right)_q\left(\frac{1-p}{2}\right)_q}{(\xi)_q(\eta)_q(\zeta)_q\, q!}\int d\Gamma\, \left[s_1^{\frac{1-m}{2}-\xi-1}(1-s_1)^{q+\xi-1}\right]\left[s_2^{\frac{1-m}{2}-1}(1-s_2)^{-\xi-1}\right]\times\\
\times\left[t_1^{\frac{1-n}{2}-\eta-1}(1-t_1)^{q+\eta-1}\right]\left[t_2^{\frac{1-n}{2}-1}(1-t_2)^{-\eta-1}\right]\times\\
\times\left[u_1^{\frac{m+n-2p+3}{2}-\zeta-1}(1-u_1)^{q+\zeta-1}\right]\left[u_2^{\frac{m+n-2p+3}{2}-1}(1-u_2)^{-\zeta-1}\right]=\\
=-\frac{(8\pi)^3\sin\left(2\pi\xi\right)\sin\left(2\pi\eta\right)\sin\left(2\pi\zeta\right)}{\xi\eta\zeta e^{i\pi(\xi+\eta+\zeta)}}\times\\
\times\sum_{q=0}^\infty\frac{\left(1-\frac{\nuk}{2}\right)_q\left(\frac{\nuk}{2}\right)_q\left(-\frac{p}{2}\right)_q\left(\frac{1-p}{2}\right)_q}{(\xi)_q(\eta)_q(\zeta)_q\, q!}\frac{(\xi)_q}{\left(\frac{1-m}{2}\right)_q}\frac{(\eta)_q}{\left(\frac{1-n}{2}\right)_q}\frac{(\zeta)_q}{\left(\frac{m+n-2p+3}{2}\right)_q}=\\
=-\frac{(8\pi)^3\sin\left(2\pi\xi\right)\sin\left(2\pi\eta\right)\sin\left(2\pi\zeta\right)}{\xi\eta\zeta e^{i\pi(\xi+\eta+\zeta)}} {}_4 F_3\left[\left. \begin{matrix} 1-\frac{\nuk}{2}& \frac{\nuk}{2} &\frac{-p}{2}& \frac{1-p}{2}\\
\frac{1-m}{2}& \frac{1-n}{2} &\frac{m+n-2p+3}{2}\end{matrix}\right|  1\right].
\end{multline}
The  additional factor is cancelled by the following constant
\begin{equation}
C_{\xi\eta \zeta}=-\frac{\xi\, \eta \, \zeta e^{i\pi(\xi+\eta+\zeta)}}{(8\pi)^3 \sin(2\pi \xi)\sin(2\pi \eta)\sin(2\pi \zeta)}.
\end{equation}
Recall that non-integer parameters $\xi,\, \eta,\, \zeta$ define the Riemann surface on which integration in carried.

Expanding even functions $f$ and $g$ in power series of $y$ and performing all the integration one can show that formula
\begin{multline}\label{evenXeven}
f(y)\ast g(y)=C_{\xi\eta \zeta}\int d\Gamma \, f\left(\sqrt{\frac{u_1 u_2}{s_1 s_2}}y\right)\sum_{p=0}^\infty \frac{i^p}{p! (u_1 u_2)^p}\left(\frac{\overleftarrow{\partial}}{\partial y_\alpha}\epsilon_{\alpha \beta} \frac{\overrightarrow{\partial}}{\partial y_\beta}\right)^p\times\\
\times\mathcal{R}(s_1,s_2,t_1,t_2,u_1,u_2)g\left(\sqrt{\frac{u_1 u_2}{t_1 t_2}}y\right)\mathcal{F}(p,\nuk,s_1,t_1,u_1)
\end{multline}
gives the same result as if product being computed with structure constants \eqref{A}. Here notation $\frac{\overleftarrow{\partial}}{\partial y_\alpha}$ means that derivative acts only on $f$ and $\frac{\overrightarrow{\partial}}{\partial y_\beta}$ acts only on $g$.

In the even case the role of the Klein operator in the decomposition
\begin{equation}
f(y,\mathcal{K})=f_0(y)+f_1(y)\mathcal{K}
\end{equation}
is trivial:
for functions even in $y$  star-product is simply the sum of products, i.e.
\begin{multline}
f(y,\mathcal{K})\ast g(y,\mathcal{K})=\left(f_0(y)+f_1(y)\mathcal{K}\right)\ast\left(g_0(y)+g_1(y)\mathcal{K}\right)=\\
=f_0(y)\ast g_0(y)+f_0(y)\ast g_1(y)\mathcal{K}+f_1(y)\ast g_0(y)\mathcal{K}+f_1(y)\ast g_1(y),
\end{multline}
where each product can be computed with the help of \eqref{evenXeven}. Note that if either function $f$ or $g$ is odd the r.h.s. of \eqref{evenXeven} vanishes because of the phase factors that appear upon integration over $s_2$ or $t_2$.

We want to obtain product for all parities in the same fashion as in even $\times$ even case. For this purpose we single out part that in final expression is obtained by differentiation $\frac{1}{p!}\left(\frac{\overleftarrow{\partial}}{\partial y_\alpha}\epsilon_{\alpha \beta} \frac{\overrightarrow{\partial}}{\partial y_\beta}\right)^p$ and then rewrite the remaining expression in the way that $m$ and $n$ appear only as Pochhammer symbols in the corresponding power series. To compensate the difference from even $\times$ even case additional factors of the form
\begin{equation}
s_1^{\mu_1}(1-s_1)^{\mu_2}s_2^{\mu_3}(1-s_2)^{\mu_4}t_1^{\mu_5}(1-t_1)^{\mu_6}t_2^{\mu_7}(1-t_2)^{\mu_8}u_1^{\mu_9}(1-u_1)^{\mu_{10}}u_2^{\mu_{11}}(1-u_2)^{\mu_{12}}
\end{equation}
are inserted. Numbers $\mu_i$ are to be defined from the transformed versions of structure constants.
\subsection{Odd $\times$ odd case}
In this section star-product of two  $y$-odd  functions is obtained. The form of the  final expression is similar to \eqref{evenXeven}.  To proceed with integral representation for structure constants \eqref{B} we rewrite this expression as
\begin{multline}\label{Bpreint}
B(m+1,n+1,p,\nuk)=\underbrace{\frac{i^p (m+1)!(n+1)!}{(m+1-p)!(n+1-p)!p!}}\Bigg[\frac{(m+1-p)(n+1-p)}{(m+1)(n+1)}F(m,n,p,-\nuk)+\\
+\frac{p(m+n-2p+3+\nuk)}{(m+1)(n+1)}F(m,n,p-1,-\nuk)+\\
+\frac{p(p-1)}{(m+1)(n+1)}\frac{m+n-2p+3+\nuk}{m+n-2p+3}\frac{m+n-2p+5+\nuk}{m+n-2p+5}F(m,n,p-2,-\nuk)\Bigg].
\end{multline}
Here the underbraced prefactor was singled out because in the final expression it is obtained by differentiation $\frac{1}{p!}\left(\frac{\overleftarrow{\partial}}{\partial y_\alpha}\epsilon_{\alpha \beta} \frac{\overrightarrow{\partial}}{\partial y_\beta}\right)^p$ and the following notation is used for brevity
\begin{equation}
F(m,n,p,\nuk)={}_4 F_3\left[\begin{matrix} 1-\frac{\nuk}{2}& \frac{\nuk}{2} &\frac{-p}{2}& \frac{1-p}{2}\\
\frac{1-m}{2}& \frac{1-n}{2} &\frac{m+n-2p+3}{2}\end{matrix}\Bigg|1\right].
\end{equation}
Let us now transform each term  in square brackets of \eqref{Bpreint} as follows.

\subsubsection{$F(m,n,p,-\nuk)$}
The prefactor before $F(m,n,p,-\nuk)$ can be rewritten as
\begin{equation}
\frac{(m+1-p)(n+1-p)}{(m+1)(n+1)}=\left(1+\frac{\frac{p}{2}}{\left(\frac{1-m}{2}-1\right)}\right)\left(1+\frac{\frac{p}{2}}{\left(\frac{1-n}{2}-1\right)}\right).
\end{equation}
Then using the definition of hypergeometric function \eqref{hyper}, Pochhammer symbol \eqref{symb} and Gamma-function identities the whole expression can be represented in the form
\begin{multline}\label{expression}
\frac{(m+1-p)(n+1-p)}{(m+1)(n+1)}F(m,n,p,-\nuk)=F(m,n,p,-\nuk)+\\
+\frac{p}{2}\sum_{q=0}^\infty\frac{\left(1+\frac{\nuk}{2}\right)_q\left(-\frac{\nuk}{2}\right)_q\left(\frac{1-p}{2}\right)_q\left(-\frac{p}{2}\right)_q}{\left(\frac{1-m}{2}-1\right)_{q+1} \left(\frac{1-n}{2}\right)_q \left(\frac{m+n-2p+3}{2}\right)_q}\frac{1}{q!}+\frac{p}{2}\sum_{q=0}^\infty\frac{\left(1+\frac{\nuk}{2}\right)_q\left(\frac{-\nuk}{2}\right)_q\left(\frac{1-p}{2}\right)_q\left(-\frac{p}{2}\right)_q}{\left(\frac{1-m}{2}\right)_{q} \left(\frac{1-n}{2}-1\right)_{q+1} \left(\frac{m+n-2p+3}{2}\right)_q}\frac{1}{q!}+\\
+\frac{p^2}{4}\sum_{q=0}^\infty\frac{\left(1+\frac{\nuk}{2}\right)_q\left(-\frac{\nuk}{2}\right)_q\left(\frac{1-p}{2}\right)_q\left(-\frac{p}{2}\right)_q}{\left(\frac{1-m}{2}-1\right)_{q+1} \left(\frac{1-n}{2}-1\right)_{q+1} \left(\frac{m+n-2p+3}{2}\right)_q}\frac{1}{q!}.
\end{multline}
Slight modification of Pochhammer symbols from even $\times$ even case like
\begin{equation}
\frac{1}{\left(\frac{1-m}{2}\right)_q}\rightarrow\frac{1}{\left(\frac{1-m}{2}-1\right)_{q+1}}
\end{equation}
in integral representation can be easily compensated  by introducing additional factors of $\frac{1-s_2}{s_2}$ or $\frac{1-t_2}{t_2}$ for
\begin{equation}
\frac{1}{\left(\frac{1-n}{2}\right)_q}\rightarrow\frac{1}{\left(\frac{1-n}{2}-1\right)_{q+1}}
\end{equation}
or both like in the last term of \eqref{expression}.

\subsubsection{$F(m,n,p-1,-\nuk)$}
Analogously part with $F(m,n,p-1,-\nuk)$ can be represented as
\begin{multline}
\frac{p(m+n-2p+3+\nuk)}{(m+1)(n+1)}F(m,n,p-1,-\nuk)=\\=\frac{\frac{p}{2}\left[\frac{m+n-2p+5}{2}-\left(1-\frac{\nuk}{2}\right)\right]}{\left(\frac{1-m}{2}-1\right)\left(\frac{1-n}{2}-1\right)}F(m,n,p-1,-\nuk).
\end{multline}
Again using definitions and identities \eqref{hyper},\eqref{symb},\eqref{GammaIdentities} it  can be rewritten as
\begin{multline}
\frac{p(m+n-2p+3+\nuk)}{(m+1)(n+1)}F(m,n,p-1,-\nuk)=\\
=\frac{p}{2}\sum_{q=0}^\infty\frac{\left(1+\frac{\nuk}{2}\right)_q\left(\frac{-\nuk}{2}\right)_q\left(\frac{2-p}{2}\right)_q\left(\frac{1-p}{2}\right)_q}{\left(\frac{1-m}{2}-1\right)_{q+1} \left(\frac{1-n}{2}-1\right)_{q+1} }\frac{1}{q!}\left(\frac{1}{\left(\frac{m+n-2p+7}{2}\right)_{q-1}}-\left(1-\frac{\nuk}{2}\right)\frac{1}{\left(\frac{m+n-2p+5}{2}\right)_q}\right).
\end{multline}
Analogously to the previous case the modified Pochhammer symbols like  $\frac{1}{\left(\frac{m+n-2p+7}{2}\right)_{q-1}}$ and $\frac{1}{\left(\frac{m+n-2p+5}{2}\right)_q}$ can be compensated introducing additional powers of  $u_1,u_2,(1-u_1)$ or $(1-u_2)$.
\subsubsection{$F(m,n,p-2,-\nuk)$}
Since the procedure is analogous we present below only the chain of trasformations
\begin{multline}
\frac{p(p-1)}{(m+1)(n+1)}\frac{m+n-2p+3+\nuk}{m+n-2p+3}\frac{m+n-2p+5+\nuk}{m+n-2p+5}F(m,n,p-2,-\nuk)=\\=\frac{\frac{p}{2}\frac{(p-1)}{2}\left(\frac{m+n-2p+3+\nuk}{2}\right)\left(\frac{m+n-2p+5+\nuk}{2}\right)}{\left(\frac{1-m}{2}-1\right)\left(\frac{1-n}{2}-1\right)\left(\frac{m+n-2p+3}{2}\right)\left(\frac{m+n-2p+5}{2}\right)}F(m,n,p-2,-\nuk).
\end{multline}
To proceed we expand the enumerator
\begin{multline}
\frac{p(p-1)}{4}\bigg[\bigg(\frac{m+n-2p+3}{2}\bigg)\bigg(\frac{m+n-2p+5}{2}\bigg)+\nuk\bigg(\frac{m+n-2p+3}{2}\bigg)+\\
+\frac{\nuk}{2}\bigg(1+\frac{\nuk}{2}\bigg)\bigg]\sum_{q=0}^\infty\frac{\left(1+\frac{\nuk}{2}\right)_q\left(\frac{-\nuk}{2}\right)_q\left(\frac{3-p}{2}\right)_q\left(\frac{2-p}{2}\right)_q}{\left(\frac{1-m}{2}-1\right)_{q+1} \left(\frac{1-n}{2}-1\right)_q \left(\frac{m+n-2p+3}{2}\right)_{q+2}}\frac{1}{q!}=\\
=\frac{p(p-1)}{4}\sum_{q=0}^\infty\frac{\left(1+\frac{\nuk}{2}\right)_q\left(\frac{-\nuk}{2}\right)_q\left(\frac{3-p}{2}\right)_q\left(\frac{2-p}{2}\right)_q}{\left(\frac{1-m}{2}-1\right)_{q+1} \left(\frac{1-n}{2}-1\right)_q }\frac{1}{q!}\Bigg[\frac{1}{\left(\frac{m+n-2p+7}{2}\right)_q}+\nuk\frac{1}{\left(\frac{m+n-2p+5}{2}\right)_{q+1}}+\\
+\frac{\nuk}{2}\bigg(1+\frac{\nuk}{2}\bigg)\frac{1}{\left(\frac{m+n-2p+3}{2}\right)_{q+2}}\Bigg]
.
\end{multline}
Now we are in a position to write product of two odd functions with the help of integration. To simplify formulas we introduce a pair of projectors
\begin{equation}
\Pi_\pm\equiv\frac{1\pm \mathcal{K}}{2}.
\end{equation}
\begin{multline}\label{oddXodd}
f(y)\ast g(y)\Pi_\pm=C_{\xi\eta\zeta}\int d\Gamma\, f\left(\sqrt{\frac{u_1 u_2}{s_1 s_2}}y\right)\sum_{p=0}^\infty \frac{i^p}{p!(u_1 u_2)^p}\left(\frac{\overleftarrow{\partial}}{\partial y_\alpha}\epsilon_{\alpha \beta} \frac{\overrightarrow{\partial}}{\partial y_\beta}\right)^p\times\\
\mathcal{R}(s_1,s_2,t_1,t_2,u_1,u_2)\frac{\sqrt{s_1s_2t_1t_2}}{u_1 u_2}\Bigg\{
\left[1-\frac{p}{2\xi}\left(\frac{1-s_2}{s_2}\right)\right]\left[1-\frac{p}{2\eta}\left(\frac{1-t_2}{t_2}\right)\right]\mathcal{F}(p,\mp\nu,s_1,t_1,u_1)-\\
-\frac{p}{2\xi\eta}\left(\frac{1-s_2}{s_2}\right)\left(\frac{1-t_2}{t_2}\right)u_1u_2\left[\frac{u_2(1+\zeta)}{1-u_2}+\left(1\mp\frac{\nu}{2}\right)\right]\mathcal{F}(p-1,\mp\nu,s_1,t_1,u_1)+\\
+\frac{p(p-1)}{4\xi\eta}\left(\frac{1-s_2}{s_2}\right)\left(\frac{1-t_2}{t_2}\right)\left[1\mp\frac{\nu}{\zeta}\left(\frac{1-u_2}{u_2}\right)\mp\frac{\nu(2+\nu)}{4(1-\zeta)\zeta}\left(\frac{1-u_2}{u_2}\right)^2\right]\times\\
\times\mathcal{F}(p-2,\mp\nu,s_1,t_1,u_1)\Bigg\}g\left(\sqrt{\frac{u_1u_2}{t_1 t_2}}y\right)\Pi_\pm.
\end{multline}
Product without projector may be obtained simply as the sum
\begin{equation}
f(y)\ast g(y)\Pi_++f(y)\ast g(y)\Pi_-=f(y)\ast g(y).
\end{equation}
Note that r.h.s. of \eqref{oddXodd} vanishes due to integration over $s_2$ or $t_2$ if either function $f$ or $g$ is even.

\subsection{Even $\times$ odd case}
Since transformation of structure constants \eqref{C} is analogous after we present the final result only
\begin{multline}
C(m,n+1,p,\nuk)=\frac{i^p m! (n+1)!}{(m-p)!(n+1-p)! p!}\Bigg\{F(m,n,p,-\nuk)+\\
+\frac{p}{2}\sum_{q=0}^\infty \frac{\left(1+\frac{\nuk}{2}\right)_q\left(-\frac{\nuk}{2}\right)_q}{\left(\frac{1-m}{2}\right)_q\left(\frac{1-n}{2}-1\right)_q q!}\Bigg[\frac{\left(\frac{1-p}{2}\right)_q\left(-\frac{p}{2}\right)_q}{\left(\frac{m+n-2p+3}{2}\right)_q}-\frac{\left(\frac{2-p}{2}\right)_q\left(\frac{1-p}{2}\right)_q}{\left(\frac{m+n-2p+5}{2}\right)_q}-\frac{\nuk}{2}\frac{\left(\frac{2-p}{2}\right)_q\left(\frac{1-p}{2}\right)_q}{\left(\frac{m+n-2p+3}{2}\right)_{q+1}}\Bigg]\Bigg\}.
\end{multline}
Star product for even function $f$ and odd function $g$ has the form
\begin{multline}\label{evenXodd}
f(y)\ast g(y)\Pi_\pm=C_{\xi\eta\zeta}\int d\Gamma\, f\left(\sqrt{\frac{u_1u_2}{s_1s_2}}y\right)\sum_{p=0}^\infty \frac{i^p}{p!(u_1 u_2)^p}\left(\frac{\overleftarrow{\partial}}{\partial y_\alpha}\epsilon_{\alpha \beta} \frac{\overrightarrow{\partial}}{\partial y_\beta}\right)^p\times\\
\mathcal{R}(s_1,s_2,t_1,t_2,u_1,u_2)\sqrt{\frac{t_1t_2}{u_1 u_2}}\Bigg\{\left[1-\frac{p}{2\eta}\left(\frac{1-t_2}{t_2}\right)\right]\mathcal{F}(p,\mp \nu,s_1,t_1,u_1)+\\
+\frac{p}{2\eta}\left(\frac{1-t_2}{t_2}\right)\left[u_1u_2\mp\frac{\nu}{2\zeta}\left(\frac{1-u_2}{u_2}\right)\right]\mathcal{F}(p-1,\mp \nu,s_1,t_1,u_1)\Bigg\}g\left(\sqrt{\frac{u_1u_2}{t_1t_2}}y\right)\Pi_\pm.
\end{multline}
R.h.s. of \eqref{evenXodd} vanishes due to integration over $s_2$ or $t_2$ if either $f$ is not even or $g$ is not odd.

\subsection{Odd $\times$ even case}
Analogously to previous section the transformed
structure constants
\begin{multline}
D(m+1,n,p,\nuk)=\frac{i^p (m+1)! n!}{(m+1-p)!(n-p)! p!}\Bigg\{F(m,n,p,\nuk)+\\
+\frac{p}{2}\sum_{q=0}^\infty \frac{\left(1-\frac{\nuk}{2}\right)_q\left(\frac{\nuk}{2}\right)_q}{\left(\frac{1-m}{2}-1\right)_q\left(\frac{1-n}{2}\right)_q q!}\Bigg[\frac{\left(\frac{1-p}{2}\right)_q\left(-\frac{p}{2}\right)_q}{\left(\frac{m+n-2p+3}{2}\right)_q}-\frac{\left(\frac{2-p}{2}\right)_q\left(\frac{1-p}{2}\right)_q}{\left(\frac{m+n-2p+5}{2}\right)_q}+\frac{\nuk}{2}\frac{\left(\frac{2-p}{2}\right)_q\left(\frac{1-p}{2}\right)_q}{\left(\frac{m+n-2p+3}{2}\right)_{q+1}}\Bigg]\Bigg\}.
\end{multline}
And product for odd and even functions
\begin{multline}\label{oddXeven}
f(y)\ast g(y)\Pi_\pm=C_{\xi\eta\zeta}\int d\Gamma\, f\left(\sqrt{\frac{u_1u_2}{s_1s_2}}y\right)\sum_{p=0}^\infty \frac{i^p}{p!(u_1 u_2)^p}\left(\frac{\overleftarrow{\partial}}{\partial y_\alpha}\epsilon_{\alpha \beta} \frac{\overrightarrow{\partial}}{\partial y_\beta}\right)^p\times\\
\mathcal{R}(s_1,s_2,t_1,t_2,u_1,u_2)\sqrt{\frac{s_1s_2}{u_1 u_2}}\Bigg\{\left[1-\frac{p}{2\xi}\left(\frac{1-s_2}{s_2}\right)\right]\mathcal{F}(p,\pm \nu,s_1,t_1,u_1)+\\
+\frac{p}{2\xi}\left(\frac{1-s_2}{s_2}\right)\left[u_1u_2\pm\frac{\nu}{2\zeta}\left(\frac{1-u_2}{u_2}\right)\right]\mathcal{F}(p-1,\pm \nu,s_1,t_1,u_1)\Bigg\}g\left(\sqrt{\frac{u_1u_2}{t_1t_2}}y\right)\Pi_\pm.
\end{multline}
R.h.s. of \eqref{oddXeven} vanishes due to integration over $s_2$ or $t_2$ if either $f$ is not odd or $g$ is not even.

\section{Full star product}
Products for different parities \eqref{evenXeven},\eqref{oddXodd},\eqref{evenXodd} and \eqref{oddXeven} schematically have the form
\begin{equation}
f(y)\ast g(y)\Pi_\pm=\int d\Gamma\, f\left(\sqrt{\frac{u_1u_2}{s_1s_2}}y\right)\mathsf{Ker}^{IJ}\left(\frac{\overleftarrow{\partial}}{\partial y_\alpha}\epsilon_{\alpha \beta} \frac{\overrightarrow{\partial}}{\partial y_\beta},s_{1,2},t_{1,2},u_{1,2}\right)g\left(\sqrt{\frac{u_1u_2}{t_1t_2}}y\right)\Pi_\pm,
\end{equation}
where $\mathsf{Ker}$ for even $\times$ even case from \eqref{evenXeven} is
\begin{multline}
\mathsf{Ker}^{EE}\left(\frac{\overleftarrow{\partial}}{\partial y_\alpha}\epsilon_{\alpha \beta} \frac{\overrightarrow{\partial}}{\partial y_\beta},s_{1,2},t_{1,2},u_{1,2}\right)=C_{\xi\eta\zeta}\sum_{p=0}^\infty \frac{i^p}{p! (u_1 u_2)^p}\left(\frac{\overleftarrow{\partial}}{\partial y_\alpha}\epsilon_{\alpha \beta} \frac{\overrightarrow{\partial}}{\partial y_\beta}\right)^p\times\\
\times\mathcal{R}(s_1,s_2,t_1,t_2,u_1,u_2)\mathcal{F}(p,\nu,s_1,t_1,u_1)
\end{multline}
And if functions do not obey certain parity requirement the integral with corresponding kernel $\mathsf{Ker}$ vanishes. Hence the product of two functions can be written as integral with the sum of kernels for all possible cases, i.e. for generic functions $f$ and $g$ the product has the form
\begin{multline}\label{fullSP}
f(y)\ast g(y)\Pi_\pm=C_{\xi\eta \zeta}\int d\Gamma\, f\left(\sqrt{\frac{u_1u_2}{s_1s_2}}y\right)\sum_{p=0}^\infty \frac{i^p}{p!(u_1 u_2)^p}\left(\frac{\overleftarrow{\partial}}{\partial y_\alpha}\epsilon_{\alpha \beta} \frac{\overrightarrow{\partial}}{\partial y_\beta}\right)^p\times\\
\mathcal{R}(s_1,s_2,t_1,t_2,u_1,u_2)\Bigg\{\mathcal{F}(p,\pm\nu,s_1,t_1,u_1)+\\
+\sqrt{\frac{t_1t_2}{u_1u_2}}\Bigg(\left[1-\frac{p}{2\eta}\left(\frac{1-t_2}{t_2}\right)\right]\mathcal{F}(p,\mp \nu,s_1,t_1,u_1)+\\
+\frac{p}{2\eta}\left(\frac{1-t_2}{t_2}\right)\left[u_1u_2\mp\frac{\nu}{2\zeta}\left(\frac{1-u_2}{u_2}\right)\right]\mathcal{F}(p-1,\mp \nu,s_1,t_1,u_1)\Bigg)+\\
+\sqrt{\frac{s_1s_2}{u_1u_2}}\Bigg(\left[1-\frac{p}{2\xi}\left(\frac{1-s_2}{s_2}\right)\right]\mathcal{F}(p,\pm \nu,s_1,t_1,u_1)+\\
+\frac{p}{2\xi}\left(\frac{1-s_2}{s_2}\right)\left[u_1u_2\pm\frac{\nu}{2\zeta}\left(\frac{1-u_2}{u_2}\right)\right]\mathcal{F}(p-1,\pm \nu,s_1,t_1,u_1)\Bigg)+\\
+\frac{\sqrt{s_1s_2t_1t_2}}{u_1u_2}\Bigg(
\left[1-\frac{p}{2\xi}\left(\frac{1-s_2}{s_2}\right)\right]\left[1-\frac{p}{2\eta}\left(\frac{1-t_2}{t_2}\right)\right]\mathcal{F}(p,\mp\nu,s_1,t_1,u_1)-\\
-\frac{p}{2\xi\eta}\left(\frac{1-s_2}{s_2}\right)\left(\frac{1-t_2}{t_2}\right)u_1u_2\left[\frac{u_2(1+\zeta)}{1-u_2}+\left(1\mp\frac{\nu}{2}\right)\right]\mathcal{F}(p-1,\mp\nu,s_1,t_1,u_1)+\\
+\frac{p(p-1)}{4\xi\eta}\left(\frac{1-s_2}{s_2}\right)\left(\frac{1-t_2}{t_2}\right)\left[1\mp\frac{\nu}{\zeta}\left(\frac{1-u_2}{u_2}\right)\mp\frac{\nu(2+\nu)}{4(1-\zeta)\zeta}\left(\frac{1-u_2}{u_2}\right)^2\right]\times\\
\times\mathcal{F}(p-2,\mp\nu,s_1,t_1,u_1)\Bigg)\Bigg\}g\left(\sqrt{\frac{u_1 u_2}{t_1 t_2}}y\right)\Pi_\pm.
\end{multline}

\section{Conclusion}
The analogue of Moyal differential star-product formula \eqref{Moyal} is obtained. It requires six additional integration parameters but in arguments of multiplied functions they appear only as certain combinations, namely $\sqrt{\frac{u_1 u_2}{s_1 s_2}}$ and $\sqrt{\frac{u_1 u_2}{t_1 t_2}}$. This  fact suggests that there should be a proper change of integration variables that decreases the number of integration parameters. The role of non-integer numbers $\xi,\eta$ and $\zeta$ is not clear at this stage, they define the Riemann surface on which integration along Pochhamer contour is carried.  Perhaps certain choice of them allows to perform a change of integration variables mentioned earlier.

As mentioned in Introduction algebra of deformed oscillators naturally appears in 3D HS gravity \cite{Prokushkin}. Two main approaches were used for computations in this theory: to realize deformed commutation relations with the doubled number of oscillators as  in the original paper \cite{Prokushkin} or to use Lone-Star product directly \cite{Perlmutter}. Even though formula \eqref{fullSP} does not look particularly promising for practical computations it can be useful for computation of products of functions
that are not just formal power series. Hopefully it can also be used to prove some general results when explicit form of functions to be multiplied is unknown.

\section*{Acknowledgments}
Author is grateful to Vyacheslav Didenko, Mikhail Vasiliev and Nikita Misuna for careful reading the paper and useful remarks. This research was supported by RFBR grant No { 20-02-00208}.

\bibliographystyle{hieeetr}

\bibliography{starbib}

\end{document}